\documentclass[aps,prl,twocolumn,showpacs,superscriptaddress,amsmath,amssymb,floatfix,altaffillsymbol]{revtex4-1}
\usepackage{graphicx}
\usepackage{ulem}
\usepackage{dcolumn}
\usepackage{bm}
\usepackage{amsmath}
\usepackage{latexsym}
\usepackage{amsfonts}
\usepackage{hyperref}
\usepackage{amssymb}
\usepackage{verbatim}
\usepackage[rightcaption]{sidecap}
\usepackage{color}
\usepackage{soul}
\usepackage{verbatim}
\usepackage[export]{adjustbox}
\usepackage{tikz}
\usetikzlibrary{shapes.geometric, arrows}
\usetikzlibrary{decorations.markings}
\usetikzlibrary{decorations.pathmorphing}
\usepackage{mathtools, nccmath, relsize}

\def\presuper#1#2%
  {\mathop{}%
   \mathopen{\vphantom{#2}}^{#1}%
   \kern-\scriptspace%
   #2}

\begin{document}
\title{Time-Stretched Spectroscopy by the Quantum Zeno Effect:
The Case of Auger Decay}
\date{\today}
\author{E. Vi\~nas Bostr\"om}
\affiliation{Lund University, Department of Physics and ETSF, PO Box 118, 221 00 Lund, Sweden}
%\affiliation{European Theoretical Spectroscopy Facility (ETSF)}
\author{M. Gisselbrecht}
\affiliation{Lund University, Department of Physics, PO Box 118, 221 00 Lund, Sweden}
\author{T.  Brage}
\affiliation{Lund University, Department of Physics, PO Box 118, 221 00 Lund, Sweden}
\author{C.-O. Almbladh}
\affiliation{Lund University, Department of Physics and ETSF, PO Box 118, 221 00 Lund, Sweden}
\author{A. Mikkelsen}
\affiliation{Lund University, Department of Physics, PO Box 118, 221 00 Lund, Sweden}
\author{C. Verdozzi}
\affiliation{Lund University, Department of Physics and ETSF, PO Box 118, 221 00 Lund, Sweden}

\begin{abstract}
A tenet of time-resolved spectroscopy is {\it faster laser pulses for shorter timescales}. Here we suggest turning this paradigm around, and slowing down the system dynamics via repeated measurements, to do spectroscopy on longer timescales. This is the principle of the quantum Zeno effect.  We exemplify our approach with the Auger process, and find that repeated measurements increase the core-hole lifetime, redistribute the kinetic energy of Auger electrons, and alter entanglement formation. We further provide an explicit experimental protocol for atomic Li, to make our proposal concrete.
\end{abstract}
\maketitle

%%%%%%%%%%%%%%%%%%%%%%%%%%%%%%%%%%%%%%%%%%%%%%%%%%%%%%%%%%%%
%%%%%%%%%%%%%%%%%%%%%%%%%%%%%%%%%%%%%%%%%%%%%%%%%%%%%%%%%%%%

Time and motion are essential entities to man's awareness of nature's changes. As such, they have been continually scrutinized by scholars, not seldom to undermine or negate their meaningfulness. A notable example was that of Zeno of Elea~\cite{Diogene}, who believed in the deceit of the ordinary perception of change and movement. In a famous conceptual paradox, he argued that an arrow should not move, since at any instant it is observed, it is at rest. 

In the Copenhagen interpretation of quantum mechanics, when a system is subject to measurement, its state is reduced. 
This leads to a quantum version of Zeno's arrow paradox~\cite{Misra77}: If an unstable system is measured upon frequently enough, it will not be able to decay. One should note that, at the quantum level, measurements may also increase the decay rate, via the so-called anti-Zeno effect (QAZE); which of the two mechanism dominates depends on the type of system and on the measurement rate~\cite{Kofman1,Kofman2,ZAZ16}.

\begin{figure}[b!]
\includegraphics[width=\columnwidth]{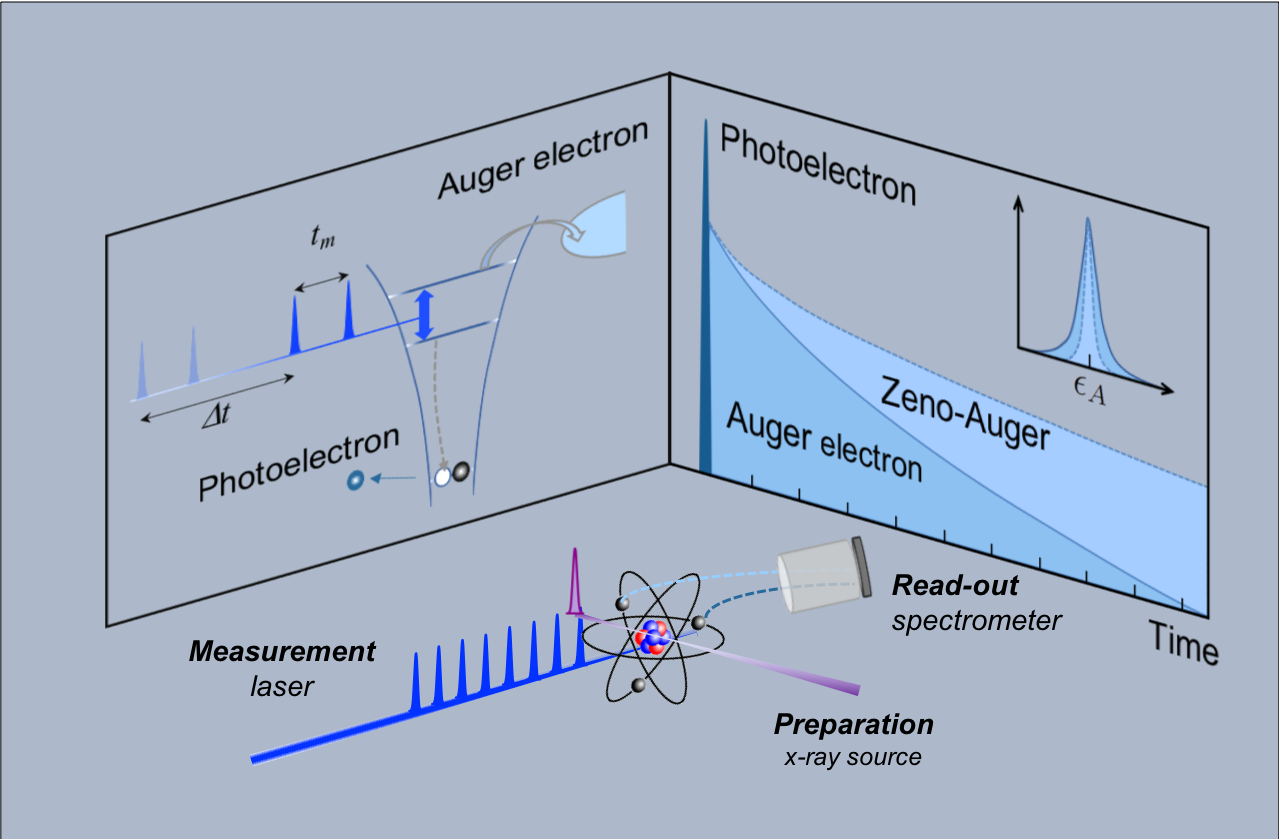}
 \caption{Pictorial rendering of an experiment on quantum Zeno effect (QZE) in Auger decay. Isolated core-hole atoms are prepared by an x-ray source arising e.g. from an accelerator based source or high harmonic generation. A pair of $\pi$-pulses separated by a time $t_m$ perform a measurement of valence electrons in the Auger-decaying states and can be repeated periodically with a delay $\Delta t$ until Auger decay occurs (left panel). Provided that the Auger decay dominates over competing processes, QZE should be observable as a slowing down of the Auger recombination time, i.e. a spectral narrowing of the Auger electron with kinetic energy, $\epsilon_A$ (right panel). Auger electron readout can be carried out by electron spectroscopy allowing hence to monitor the decay either in the energy domain (regardless of the temporal characteristics of the x-ray source), or in the time domain with attosecond pulses and a third laser field (not shown).}
 \label{fig:Auger}
\end{figure}

The quantum Zeno effect (QZE) has been realized in the laser-induced dynamics of two-level ions~\cite{Cook88,Itano90}, and in the decay of ultracold atomic gases~\cite{Fischer01,Patil15}. However, it has not yet been directly observed in natural decay processes. 
Here we take a step in bridging this gap, by proposing a protocol to measure the QZE in Auger decaying atoms (see Fig.~\ref{fig:Auger}). The Auger decay is a fundamental atomic process~\cite{Auger,Meitner,AugerMeitner} by which an inner shell vacancy (a core hole) relaxes by emitting a secondary electron. Due to the short lifetime of the core hole and the non-local nature of the interactions involved, theoretical modeling of the Auger process is challenging and, until recently, mostly performed in the energy domain~\cite{Chattarji,Weightman,Gunnarsson80,COA,Diehl,Verdozzi,Kolorenc} (see however~\cite{COA,Covito,Smirnova}). Yet, due to progress in ultrafast spectroscopy, real-time studies of the Auger decay are coming of age~\cite{Drescher,Nikolopoulos,Bonitz,Pazourek,Rohringer}, to e.g. probe photo-induced electron correlations at the few fs timescale~\cite{Schuler} or to use core-hole lifetimes as a clock for timing atomic processes~\cite{AugerClock}. 

In this Letter, we i) introduce a measurement protocol to induce QZE in atomic Auger decay,
and we demonstrate it by real-time simulations. Specifically, a train of $\pi$-pulses periodically drives
a transition from the Auger decaying state to a more stable level (see Fig.~\ref{fig:Auger}).  By increasing the pulse intensity and the repetition rate, the QZE is enhanced. 
We then ii) explicitly consider a Li atom and a hollow Li$^+$ ion, finding an increased lifetime due to QZE that should be clearly {\it experimentally detectable}. Via our simulations, we also iii) gauge the range of experimentally controllable parameters where the QZE should be observable in atomic Auger decay. Finally, we iv) show how the QZE influences the Auger lineshape and the formation of entangled continuum states. 

Our work thus provides a proof-of-concept of a novel, general notion of spectroscopy  
in systems with dynamics slowed down via the QZE.

%%%%%%%%%%%%%%%%%%%%%%%%%%%%%%%%%%%%%%%%%%%%%%%%%%%%%%%%%%%%
%%%%%%%%%%%%%%%%%%%%%%%%%%%%%%%%%%%%%%%%%%%%%%%%%%%%%%%%%%%%

{\it Atomic system and external fields.-} 
We consider a model atomic system, where a core electron has already been ejected and does not interact with the remaining system (the sudden ejection limit \cite{Hedin69}). The atom has two spinless electrons that enter the Auger decay, and is exposed to a classical time-dependent light-field $E(t)$ treated in the dipole approximation ($t$ labels time). Our choice of a spinless model is computationally convenient, while fully retaining the essential aspects of Auger physics compared to the spinful case \cite{Gunnarsson80}. The atom is modeled in terms of four atomic orbitals, $|c\rangle$, $|v_1\rangle$, $|v_2\rangle$ and $|v_3\rangle$, and the continuum orbitals $|\epsilon_k\rangle$ grouped in two regions $\mathcal{S}$ and $\mathcal{P}$ corresponding respectively to states with $s$- and $p$-symmetry.

The dynamics of the system is determined by the effective Hamiltonian $H(t) = H_0 + H_1(t)$, where 
\begin{align}\label{eq:ham}
H_0 &= E_1 |1\rangle\langle 1| + E_2 |2\rangle\langle 2| + \sum_{k\in \mathcal{S}} E_k |k\rangle\langle k| + \sum_{k\in \mathcal{P}} E_k |k\rangle\langle k| \nonumber \\
     &+ \sum_{k\in \mathcal{S}} M_k (|k\rangle\langle 1| + \text{H.c.}) + \sum_{k\in \mathcal{P}} M_k (|k\rangle\langle 2| + \text{H.c.}).
\end{align}
The Hamiltonian is here expressed in the two-particle states $|1\rangle = |v_1v_2\rangle$ and $|2\rangle = |v_1v_3\rangle$, which decay with lifetimes $\tau_1$ and $\tau_2$ into the states $|k\rangle = |\epsilon_kc\rangle$. The lifetimes are set by the matrix elements $M_k$ (related to the Auger matrix elements), for which we use the Fano approximation $M_k = M_1$ ($M_k = M_2$) for $k \in \mathcal{S}$ ($k \in \mathcal{P}$). Given a density of states $\rho(\epsilon)$ for the continuum states, this gives a one-to-one mapping $\tau_i \leftrightarrow M_i$ for $i=1,2$. For the system considered here, the effective Hamiltonian can be derived from a fundamental many-body Hamiltonian expressed in terms of single-particle orbitals, as described in the Supplemental Material (SM). It gives an exact description of the dynamics starting from the initial state $|\Psi_0\rangle = |v_1v_2\rangle = |1\rangle$.

In our numerical simulations, the continuum states span a finite energy interval centered at the Auger energy $\epsilon_A^{1(2)} = E_{1(2)}-\epsilon_c$ for region $\mathcal{S}$ ($\mathcal{P}$), and are distributed according to $\rho(\epsilon) \propto 1/\sqrt{\epsilon}$. We have checked that the results are insensitive to the change $\rho(\epsilon) \propto \epsilon^{(n-2)/2}$ for $n=1$, 2 and 3.

The interaction between the atom and the external measurement field is given by $H_1(t) = \Omega f(t)\sin(\omega t) (|2\rangle\langle 1| + |1\rangle\langle 2|)$, where $f(t)$ is an envelope function such that $f(t) = 0$ for $t<0$, and $\Omega$ is the Rabi frequency of the transition. We assume that the laser frequency $\omega$ is in resonance with the transition $|1\rangle \leftrightarrow |2\rangle$, and further that $\hbar\omega$ is smaller than the system's ionization potential.

We solved exactly the Schr\"odinger equation $i\partial_t|\psi(t)\rangle = H(t)|\psi(t)\rangle$ with the Lanczos algorithm \cite{ParkLight} to obtain the populations $n_c(t)$, $n_{v_1}(t)$, $n_{v_2}(t)$ and $n_{v_3}(t)$ of the atomic orbitals and so monitor the Auger decay in time. The Auger lineshape was calculated via the time-dependent populations $\mathcal{A}(\epsilon_k,t)= n_k(t)$ of the continuum states. Because of the high kinetic energy of the Auger electron, no reabsorption occurs (as also verified numerically).

%%%%%%%%%%%%%%%%%%%%%%%%%%%%%%%%%%%%%%%%%%%%%%%%%%%%%%%%%%%%
%%%%%%%%%%%%%%%%%%%%%%%%%%%%%%%%%%%%%%%%%%%%%%%%%%%%%%%%%%%%

\begin{figure*}[t]
 \includegraphics[width=\textwidth]{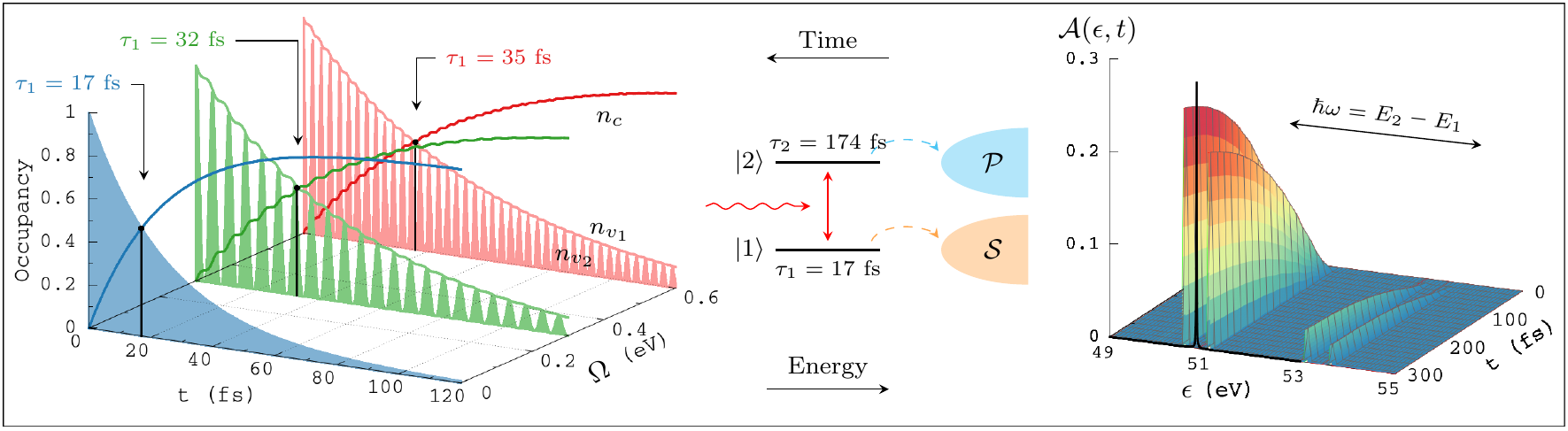}
 \caption{Auger decay and quantum Zeno effect with measurement time $t_m = 0.32$ fs. Left: electron orbital occupations $n_c$, $n_{v_1}$ and $n_{v_2}$ in a model description of Li as a function of time, for field intensities $I = 0$, $5.1$ and $20.4$ TW/cm$^2$ (blue, green and red curves, respectively) pertaining to Rabi frequencies $\hbar\Omega = 0$, $0.3$ and $0.6$ eV. The black lines indicate the lifetime $\tau_1$ of the core level, and the correspondence between orbitals and curves for $\hbar\Omega=0.6$ eV applies to all intensities. 
 Center: Schematics of system and driving field.
 Right: Occupation $\mathcal{A}(\epsilon_k,t)$ in the continuum levels $|\epsilon_k\rangle$ as a function of time and energy for $I = 20.4$ TW/cm$^2$. The long-time limit spectral line for $I = 0$ is shown in black.} 
 \label{fig:Auger_li}
\end{figure*}

{\it Quantum Zeno protocol for Auger decay.-}
To hinder the Auger process in time via the QZE, we need to periodically bring the system back to its initial state. To ``freeze'' the decay, the time $\Delta t$ between return events (measurements)  should be small compared to $\tau_1$ \cite{Pascazio}.

To this end, we suggest the following protocol~\cite{inspired,Nori13}: At time $t=0$ we send in a square pulse of duration $t_\pi = \pi/\Omega$ (a so-called $\pi$-pulse), after which the probability $P_2$ of finding the system in state $|2\rangle$ is given by $P_2(t_\pi) = P_1(0)\Omega^2/(\Omega^2 + \delta^2)$. Here $\delta = \omega - \Delta$ is the detuning from resonance, $\Delta = E_2 -E_1$ the energy separation of the states, and $P_1(0)$ the probability that the atom is initially in state $|1\rangle$. 
At this point we wait a time $t_m$, after which another $\pi$-pulse transfers the system back with probability $P_1(t_\pi+t_m+t_\pi) = P_1(0)\left[\Omega^2/(\Omega^2 + \delta^2)\right]^2$. For no detuning ($\delta = 0$), the final probability would be $P_1(2t_\pi+t_m) = P_1(0)$, and the system would return to its original state, where the Auger decay can take place. 
The whole procedure could be seen as a projective measurement: the wave ``collapses'' into  state $|1\rangle$, but only if the Auger decay has not occurred yet. However, such visualization is unnecessary, since i) our measurement is not instantaneous and ii) we can instead measure the dipole radiation induced by the oscillations $|1\rangle \leftrightarrow |2\rangle$.

{\it QZE-vs-QAZE and protocol parameters.-} 
As noted above, our protocol relies on finite time measurements. Hence, it differs from a QZE derived from projective measurements performed at an interval $\Delta t$, giving a survival probability after $N$ measurements $P(N\Delta t) = P(\Delta t)^N$. We can never take the limit $\Delta t \to 0$ and $N \to \infty$, since for us $\Delta t$ is bounded from below. We have checked numerically that including projective measurements at the end of each cycle has marginal effect for our protocol, and hence neglect them in the following. We also find that for $\tau_1 < \tau_2$, varying the time between measurements always gives a transition from unperturbed decay to QZE, with no intermediate QAZE. However, for $\tau_1 > \tau_2$, we instead find a QAZE. In contrast to projective schemes, where the  QZE to QAZE transition is a function of measurement frequency, our protocol finds it only depends on {\it system parameters} (for a full discussion see SM). Since we are interested in slowing down the Auger decay, we henceforth focus on the Zeno regime, i.e. $\tau_1 < \tau_2$.

The physical parameters suitable for the QZE protocol in the case of the Auger decay are constrained by the following observations: i) The measurement time must be significantly shorter than the Auger lifetime $\tau_1$~\cite{SM1}; since $t_\pi = \pi/\Omega$, this corresponds to having $\Omega^{-1} \ll \tau_1$. In principle, this is is accomplished by increasing the field strength $\mathcal{E}$, because $\Omega= \mathcal{E}d$. ii) The frequency of the laser cannot be larger than the ionization potential; even so, for high intensities the multiphoton ionization (MPI) rate also becomes of importance. For the systems considered below, we have used the PPT model to make sure MPI is negligible.~\cite{SM2} iii) To have efficient population transfer between states $|1\rangle$ and $|2\rangle$, we ideally need to be in the weak coupling regime $\Omega \ll \omega$. Can the conditions above be met during the Auger decay of a real atom? 
As shown next, the QZE turns out to be clearly {\it observable} for {\it realistic} systems, either in the time-domain~\cite{Drescher} or as a narrowing of the Auger spectral linewidth.

%%%%%%%%%%%%%%%%%%%%%%%%%%%%%%%%%%%%%%%%%%%%%%%%%%%%%%%%%%%%
%%%%%%%%%%%%%%%%%%%%%%%%%%%%%%%%%%%%%%%%%%%%%%%%%%%%%%%%%%%%

{\it Auger decay in Lithium, with and without QZE.-} 
We consider the Li atom and associate the atomic configurations $|1s(2s^2 \presuper{1}{S})\presuper{2}{S}^e\rangle$ and $|1s(2s2p \presuper{3}{P})\presuper{2}{P}^o\rangle$ with the states $|1\rangle$ and $|2\rangle$ of our model. Since the $1s$ electron is frozen during the decay, the problem corresponds to an effective two-particle system that can be modeled via the effective Hamiltonian. The configurations $1s2s^2$ and $1s2s2p$ have respective lifetimes of $\tau_1 = 17.6$ fs and $\tau_2 = 174$ fs, dominated by Auger decay~\cite{Verbockhaven}, and the transition between the states is optically accessible by resonant driving with a field of frequency $\hbar\omega = 2.5$ eV~\cite{SM3}.

The decay dynamics of Li is shown in the left panel of Fig.~\ref{fig:Auger_li}. With no external field the Auger transition from the state $1s2s^2$ happens with a lifetime $\tau_1 = 17.0$ fs. Driving the transition $1s2s^2 \leftrightarrow 1s2s2p$ with a field of intensity $I=5.1$ TW/cm$^2$ and a measurement time $t_m = 0.32$ fs, the lifetime of the state $1s2s^2$ is extended to $\tau_1 = 32.7$ fs, and further to $\tau_1 = 35.3$ fs by increasing $I$ to $I = 20.4$ TW/cm$^2$.

We can also analyze the decay via the occupation $\mathcal{A}(\epsilon_k, t)$ of the electrons emitted into the continuum, and detect how the Auger spectral peaks arise in time (Fig.~\ref{fig:Auger_li}, right panel). Without external field, $\mathcal{A}$ has a single peak in the long-time limit. Conversely, when measurements are performed (nonzero field), there are two peaks, resulting from the decay of the $1s2s^2$ and $1s2s2p$ levels. Each peak is split by the dynamical Stark effect into two subpeaks separated by $\Delta\epsilon = \hbar\Omega$, for a total of four peaks. 

As further evidence that the QZE is measurable, we also considered hollow Li$^+$ (see SM), finding that the lifetime of the configuration $|2s^2 \presuper{1}{S}_0\rangle$ is extended from $\tau_1 = 3.3$ to $\tau_1 = 4.7$ fs by driving the transition to $|2s2p \presuper{1}{P}_1\rangle$. Overall, the Li and Li$^+$ results are a clear proof of concept that it is possible to stretch (and slow down) the Auger decay via QZE. Although we found only a slowing down of the Auger transition, the effect is large enough to be clearly measurable either in the time-domain~\cite{Drescher} or as a narrowing of the spectral linewidth. Of relevance to possible experimental realizations, we checked that a similar QZE is found by replacing the pulse-train with continuous radiation, corresponding to the limits $t_m \to 0$ and $\Delta t \to 2t_\pi$ (see SM).

%%%%%%%%%%%%%%%%%%%%%%%%%%%%%%%%%%%%%%%%%%%%%%%%%%%%%%%%%%%%
%%%%%%%%%%%%%%%%%%%%%%%%%%%%%%%%%%%%%%%%%%%%%%%%%%%%%%%%%%%%

\begin{figure}[t!]
 \includegraphics[width=\columnwidth]{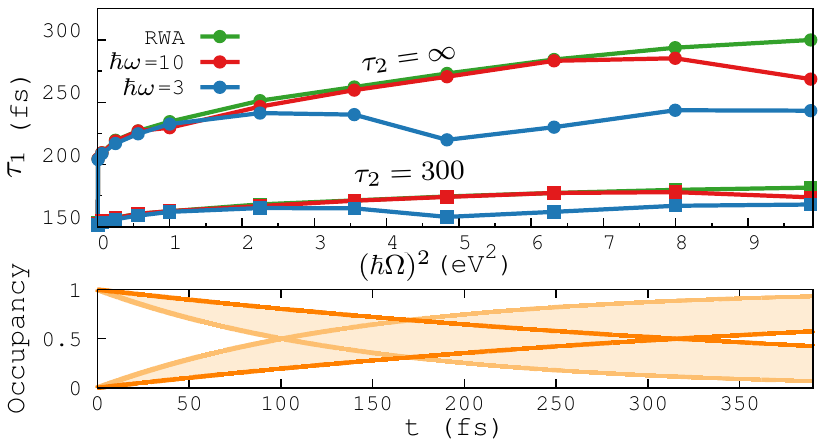}
 \caption{Parameter dependence of the QZE protocol. Top panel: Effective lifetime as a function of the squared Rabi frequency $(\hbar\Omega)^2 \propto I$. The circles show a system with unperturbed lifetimes $\tau_1 = 100$ fs and $\tau_2 = \infty$, from RWA (green), full field dynamics for $\hbar\omega = 10 $ eV (red) and $\hbar\omega = 3$ eV (blue). The lines are a guide to the eye.
 Squares: same as circles, but with $\tau_2 = 300$ fs. Bottom panel: Orbital occupations $n_c$ and $n_1$ as function of time for $\tau_2 = \infty$ and within RWA. The boundaries of such areas are the cases with no field and $\hbar\Omega = \sqrt{10}$ eV. \label{fig:model}}
\end{figure}

{\it QZE-vs-Auger trends.-} 
We now assess the role of the lifetimes $\tau_1$ and $\tau_2$, the level spacing $\hbar\omega = E_2-E_1$, and the Rabi frequency $\Omega$ ($\Omega^2$ is proportional to the field intensity $I$) in QZE, and start by considering $\tau_{2} = \infty$. If $\omega \gg \Omega$ our protocol permits to extend $\tau_1$ to many times its unperturbed value. Interestingly, this is the regime where a rotating wave approximation (RWA)~\cite{RWA} treatment and the full field give the same dynamics (see Fig.~\ref{fig:model}). In contrast, when $\omega \approx \Omega$, the RWA overestimates the increase of the lifetime. However, even for these parameters the lifetime can be extended enough for the effect to be clearly measurable. For $\tau_2$ finite but larger than $\tau_1$ similar results are observed: For $\omega \gg \Omega$ it is possible to significantly extend $\tau_1$, but now with an expected upper bound $\tau_2$ (see SM). For $\Omega \approx \omega$, i.e. for strong fields, we again find that the RWA overestimates the effective lifetime.

As shown in Fig.~\ref{fig:model}, our protocol performs best for weak coupling $\Omega \ll \omega$, where the effective lifetime can be significantly enhanced. This also clarifies why we don't observe a full halt of the decay in  Li and Li$^+$: 
the reason is a combination of the fast decay times ($\tau_{1} \approx 17.3$ fs in Li and $\tau_{1} \approx 3.4$ fs in Li$^+$) and the small transition energies ($\hbar\omega = 2.5$ eV in Li and $\hbar\omega = 4.1$ eV in Li$^+$). The short lifetimes require a high intensity for $\Omega^{-1}$ to be comparable with $\tau_1$, but the high intensities make $\Omega$ comparable to $\omega$. This prevents $\tau_1$ to be extended beyond its value for $\Omega \approx \omega$. Although measurable already for Li and Li$^+$, the QZE should be more pronounced in systems with longer lifetimes and greater transition energies.
In summary, the transition energy and the lifetimes of the two levels, all have a great influence on the occurrence of the QZE in the Auger decay.
At least in the weak intensity limit, to maintain the applicability of RWA, one could use,
instead of square pulses, pulses where the intensity and frequency can be changed 
as a function of time. This is known to improve population transfer in e.g.
NMR spectroscopy \cite{Silver} and quantum information \cite{Roos04}.

%%%%%%%%%%%%%%%%%%%%%%%%%%%%%%%%%%%%%%%%%%%%%%%%%%%%%%%%%%%%
%%%%%%%%%%%%%%%%%%%%%%%%%%%%%%%%%%%%%%%%%%%%%%%%%%%%%%%%%%%%

\begin{figure}[t!]
 \includegraphics[width=\columnwidth]{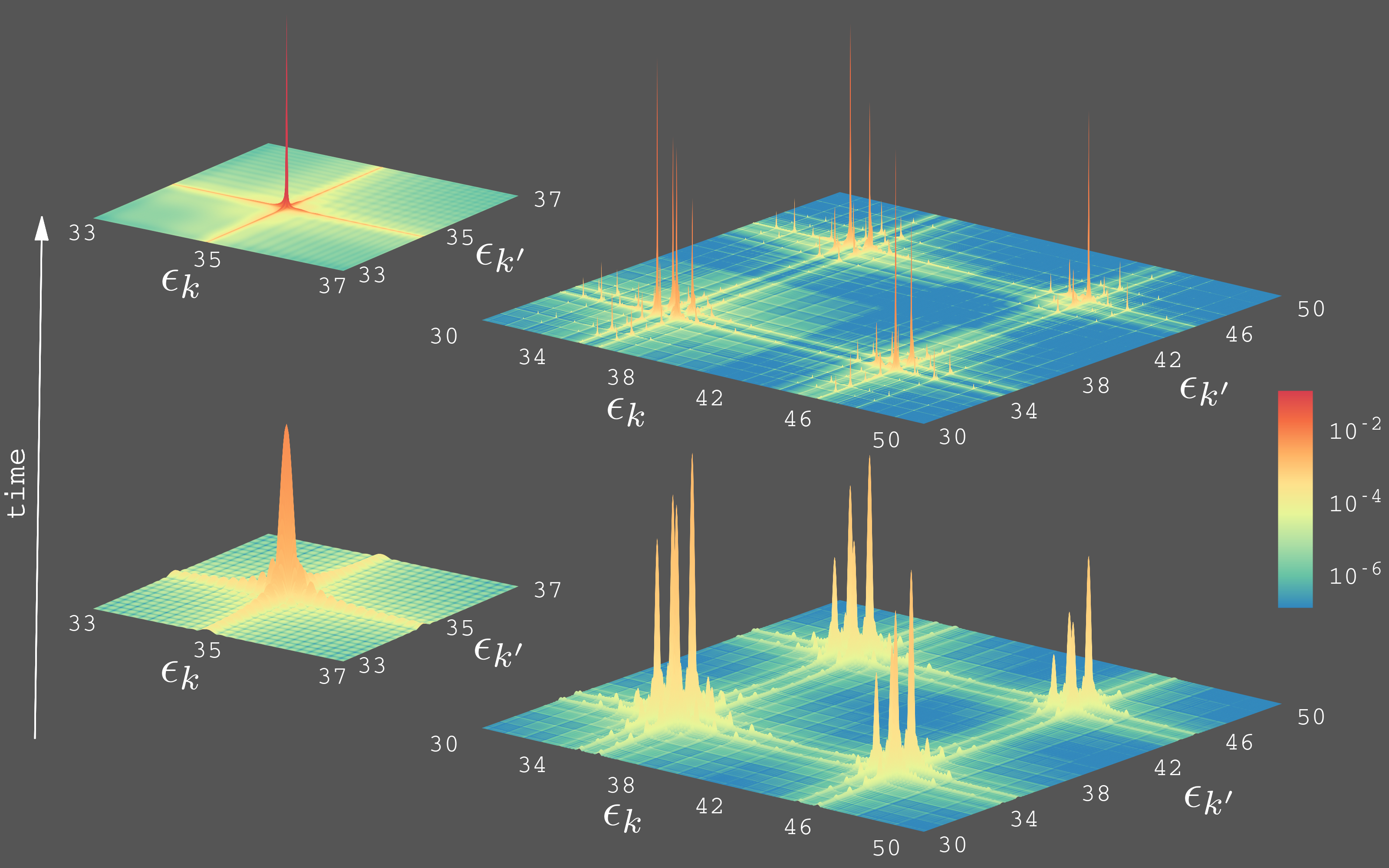}
 \caption{Concurrence matrix $\mathcal{C}$. Left: snapshots of $\mathcal{C}_{\epsilon_k \epsilon_{k'}}$ at $t = 12$ fs and $380$ fs, for no-field Auger decay with $\hbar\omega = 10$ eV, $\tau_1 = 100$ fs and $\tau_2 = 300$ fs. Right: same, but for field intensity $I = 210$ TW/cm$^2$ and measurement time $t_m = 0.32$ fs. The $\mathcal{C}$ profiles are shown on linear vertical scales, and colored according to their value on the log-scale color bar.\label{fig:entang}}
\end{figure}

{\it Auger decay, QZE, and entanglement.-} 
Having in focus the interplay of Auger decay and QZE, our treatment does not keep track of the primary (core) photo-electron. Thus, a description of the entanglement between photo- and Auger electrons, as measured in coincidence experiments~\cite{Haak,Schoffler,MG14} is not viable \cite{explain}. However, we can still explore how QZE affects entanglement formation in the Auger continuum. This is interesting in itself, as a clear test of the necessity of a coherent description of competing decay channels for QZE in Auger decay.
In general, the choice of an entanglement measure is dictated by the situation at hand. In our case, with the photo-electron not treated explicitly, we use mode concurrence~\cite{Zanardi}. In Fig.~\ref{fig:entang} we show the concurrence matrix $\mathcal{C}_{\epsilon_k \epsilon_{k'}}$ without and with external fields. With no field there is a single Auger peak at $\epsilon_A \approx 35$ eV, and there is concurrence between this state and all other continuum states. With the field there are two peaks at $\epsilon_A \approx 35$ eV and $45$ eV that are split due to the Stark effect. In this case there is concurrence within each continuum, but also between the different continua, suggesting an interesting interplay between QZE \cite{Pascazio,Maniscalco}, Auger decay, coherence and entanglement formation among different Auger channels in the continuum \cite{futurework}.

{\it Conclusions and outlook.-} 
We showed that the Auger lifetime of an atom can be increased due to the quantum Zeno effect. To this end, we proposed a protocol based on periodic driving of a bound-bound transition during the Auger decay, either with pulsed or continuous radiation (for bound-continuum transitions see SM~\cite{SM4}). As concrete example we considered the Li atom, showing that the physical parameter values to be used are within experimental reach.

The Auger decay is an important, fast and natural de-excitation
process in atoms, and this is why we chose it for a proof-of-concept of our proposal
of a "time-stretched spectroscopy". More in general, it should be possible to do the same with other natural (or not) decay phenomena. More precisely, one can envisage pump-probe experiments where, after an initial Hamiltonian quench, the ensuing relaxation dynamics can be studied on artificially longer timescales, thanks to repeated measurements which induce the quantum Zeno effect, and thus slow down the system's time evolution.

%%%%%%%%%%%%%%%%%%%%%%%%%%%%%%%%%%%%%%%%%%%%%%%%%%%%%%%%%%%%
%%%%%%%%%%%%%%%%%%%%%%%%%%%%%%%%%%%%%%%%%%%%%%%%%%%%%%%%%%%%

\begin{acknowledgments}
We acknowledge S. Maniscalco, S. Ristinmaa S\"orensen and  S. Carlstr\"om for useful discussions. E. V. B. acknowledges support from Crafoordska stiftelsen. M.G. acknowledges support by the Lund Attosecond Science Center. T.B. , A. M. and C. V. were supported by the Swedish Research Council.
\end{acknowledgments}


\begin{thebibliography}{[10]}
\bibitem{futurework} A complete discussion of this aspect, inclusive of the role of the photoelectron, is left to future work.
\end{thebibliography}
\end{document}